\documentclass{aa}
\usepackage{graphicx}
\usepackage{txfonts}
\def\lya{Ly$\alpha$ }
\def\lbgn{$LBG_N$ }
\def\lbgl{$LBG_L$ }
\begin{document}
\title{Physical and morphological properties of z$\sim 3$ LBGs: dependence on Ly$\alpha$ line emission  }
%   \subtitle{I. Overviewing the $\kappa$-mechanism}

   \author{   L.Pentericci\inst{1}
             \and 
          A. Grazian\inst{1} \and  C. Scarlata\inst{2} \and A. Fontana\inst{1} \   \and M. Castellano\inst{1}  \and E. Giallongo\inst{1} \and E. Vanzella\inst{3} }
\offprints{Laura Pentericci penteric@oa-roma.inaf.it}

   \institute{INAF - Osservatorio Astronomico di Roma, Via Frascati 33,
I--00040, Monte Porzio Catone, Italy
  \and Spitzer Science Center, Pasadena, CA, USA
   \and INAF-Osservatorio Astronomico di Trieste, via G. B. Tiepolo 11, 40131 Trieste, Italy}
   \date{}

\abstract{}{We investigate the physical and morphological 
properties of LBGs at redshift $\sim$2.5 to $\sim$3.5, to  determine if and how they depend on the nature and strength of the \lya emission. } {We selected U-dropout galaxies from the z-detected GOODS-MUSIC catalog, by adapting the classical Lyman Break criteria on the GOODS filter set. We kept only those galaxies with spectroscopic confirmation, mainly from VIMOS and FORS public observations. 
Using the  full multi-wavelength 14-bands 
information (U to IRAC), we determined the physical properties of the galaxies, 
through a standard spectral energy distribution fitting procedure with the updated Charlot \& Bruzual (2009) templates.
We also added other relevant observations of the GOODS field, i.e. 
the 24$\mu m$ observations from Spitzer/MIPS and the 2 MSec Chandra X-ray 
observations. Finally, using non parametric diagnostics (Gini, Concentration, Asymmetry, $M_{20}$ and ellipticity), we characterized the rest-frame UV morphologies of the galaxies. We then analyzed how these physical and morphological properties correlate with the presence of the \lya\ emission line in the optical spectra.}{We find that, unlike at higher redshift, the dependence of physical properties on the Ly$\alpha$ line is milder: galaxies without Ly$\alpha$ in emission tend to be more massive and dustier than the rest of the sample, but all other parameters, ages, star formation rates (SFR), X-ray emission as well as UV morphology do not depend strongly on the presence of the \lya\ emission. A simple scenario where all LBGs have intrinsically high Ly$\alpha$ emission, but where dust and neutral hydrogen content (which shape the final appearance of the Ly$\alpha$) depend on the mass of the galaxies, is able to reproduce the majority of the observed properties at z$\sim$3. Some modification might be needed to account for the observed evolution of these properties with cosmic epoch, which is also discussed.}{}
\maketitle
\section{Introduction}
A long debated issue concerns the relation between Lyman break galaxies (LBGs), i.e. star-forming  galaxies selected from the presence of the
Lyman break in their spectral energy distribution  (e.g. Steidel et al. 1996),  and Ly$\alpha$ emitters (LAEs), i.e. star-forming galaxies selected via the presence of a strong \lya\ emission line, through deep 
 narrow-band observations  (e.g Cowie \& Hu 1998, Hu et al. 1998, Rhoads \& Malhotra 2001).
Both these techniques have led to the discovery of large number of galaxies at increasingly high redshift (e.g. Ouchi et al. 2005, Venemans et al. 2007, Stanway et al. 2007, Ota et al. 2008), with the current record holder being a LAE at z=6.96 (Iye et al. 2006).
The nature of the high redshift galaxies selected through the \lya\ emission line and the  link with the star-forming population selected via the Lyman--Break is still not clear (e.g., \cite{gaw09}). This relation has an important  implication 
in our interpretation of the very high redshift Universe where, due to current 
instrumental limitations, it becomes progressively  easier to spectroscopically confirm 
 only the \lya emitters (e.g. Dow-Hygelund et al. 2007). 

On one hand the \lya\ bright phase could represent a stochastic event in the life of any galaxy (as in 
the duty cycle model e.g. Nagamine et al. 2008); alternatively it  could be related to some specific
 physical property of the galaxies, such as the (young) age, 
the dust or gas content. 
\\
 In principle, the Ly$\alpha$ line can be used to probe star-formation rates, clustering properties (Kovac et al. 2007), and even for cosmological applications such as re-ionization studies (e.g. Dijkstra et al. 2007, Dayal et al. 2008). In practice, this effort is far from trivial because of the above stated complications. 
Understanding what determines the presence of the \lya 
emission line is therefore fundamental to interpret the 
bias introduced by the observational selection techniques.

A comparison between the physical and morphological
properties of LBGs and LAEs is necessary but not straightforward. 
Because of the different selection techniques,
LAEs tend to be much fainter than LBGs and therefore modelling their
 spectral energy distributions (SEDs) to constrain the relevant physical properties 
has always been a difficult 
task (see Gawiser 2009 for a review). Most studies rely on stacked photometry 
and are able to derive only average properties (e.g. Finkelstein et al. 2007, 
Nilsson et al. 2007, Lai et al. 2008).
Furthermore until recently,  most  LAEs samples, 
lacked (deep) data in 
the  mid-IR range which are crucial to constrain stellar mass at redshift $>$ 4, 
where the 4000 \AA\ break is shifted beyond the K-band.
A good coverage of 
the  near-IR was also often missing, and this is important 
to  reduce the model degeneracies (e.g old/dust-free vs young/dusty population).
For this reason, we took a slightly different approach and decided to study 
the properties of Ly$\alpha$ emitting galaxies selected as LBGs.
In this way, we select  galaxies that are bright enough in the continuum   to be detected at other wavelengths (so their SEDs can be modelled) and show 
 Ly$\alpha$ in emission: we can then study the  dependence of physical properties on the emission line strength and characteristics, on individual galaxies, rather than relying on the stacking technique.
Clearly this is only possible with a survey that has both deep multi-band data on a relatively large area, as well as  excellent spectroscopic coverage.
The GOODS survey (e.g., Dickinson et al. 2004, Giavalisco et al. 2004) has all these properties: in particular we took advantage of the excellent observations 
in the near-IR, which cover the rest-frame region around the 4000 \AA\ Balmer break, reducing  some of the model degeneracies.  Furthermore, the deep IRAC data allow the determination  of stellar masses  with great  accuracy (e.g. Fontana et al. 2006).
\\
In previous works, we have compared the properties of LBGs with and without line emission 
for a relatively small sample of z$\sim$ 4 galaxies 
(Pentericci et al. 2007, hereafter P07),  
 and then studied the properties of a more numerous sample  of Ly$\alpha$ emitting LBGs 
in the redshift range  3 to 6 (Pentericci et al. 2009, hereafter P09).
In this paper we study the properties of LBGs at redshift $\sim$2.5-3.5, 
the so-called U--dropouts, initially  selected from their continuum 
and color properties and with follow-up spectroscopic confirmation: 
the lower redshift allows us to assemble a sample much larger than in our 
previous work and study the trends in a statistically
 more significant way.  At the same time thanks to 
the brighter average magnitude of the galaxies  we can also attempt 
a  morphological  analysis, which was not possible at higher z.
\\
As in previous work, we use the  GOODS--MUSIC catalog (Grazian et al. 2006) 
in its  revised and updated version (Santini et al. 2009). In particular the 
new cataloger contains  24 micron data from Spitzer/MIPS observations and a revised and more accurate IRAC photometry. New spectra from public surveys (Vanzella et al. 2009; Popesso et al. 2009) were also added.
We also include an analysis of the deep  X-ray  data available for the field i.e. 
the 2 Ms Chandra exposure  (Luo et al. 2008) that allows us to study the star 
formation and/or AGN content in an independent way. 
\\
All magnitudes are in the AB system (except where otherwise stated) and we adopt the
$\Lambda$-CDM concordance cosmological model ($H_0=70$, $\Omega_M=0.3$ and
$\Omega_{\Lambda}=0.7$).
\section{The sample }
To select z$\sim 3 $ galaxies we 
 have adapted the usual Lyman break technique to the filters adopted  in 
the GOODS survey, which is different from the classical UGR set used by 
Steidel et al. 2003 (see also Giavalisco et al. 2004). 
The redshift $\sim 3$ objects are selected as U--dropouts according to
the following cuts:

$$\displaystyle -0.2 \le V-I \le 0.35$$ 	 
  	  	$$\displaystyle U-V \ge 0.75(V-I)+1.15$$
which are effective at $2.4\le z\le 3.5$.
These cuts were determined by Grazian et al. (2007) 
using our multicolor GOODS-MUSIC database, and from  a comparison
to the photometric redshift distribution were adjusted  to maximize the completeness and  minimize the 
number of interlopers at redshift outside the above range.
In that work, we also verified that these criteria are almost 
(although not entirely) equivalent to the original LBG criteria, 
by reproducing the synthetic UGR magnitudes for
the galaxies.
However several differences remain, beside the difference in the filter set 
and the color selection: first of all
the LBGs were traditionally selected in the R-band, which is lacking in the GOODS survey, and not in the z-band, as in this work; furthermore LBGs were originally relatively bright galaxies, since the historical criterion used was $R\le 25.5$, while in this work the selection is pushed till the nominal completeness of the GOODS-MUSIC sample (z-band  magnitude limit of 26.0).  
All these result in some differences in the final sample properties, mainly 
a redshift distribution  of galaxies that is slightly wider than in Steidel 
et al. (2003) sample and possibly a slightly  bluer average color 
(see also Section 4.1 for this point).
 Objects flagged  as AGNs (on the basis  of their X-ray emission,and/or mid-IR excess and/or  spectral properties) in the GOODS-MUSIC catalog were not included in the analysis.
\begin{figure}
\includegraphics[width=9cm,clip=]{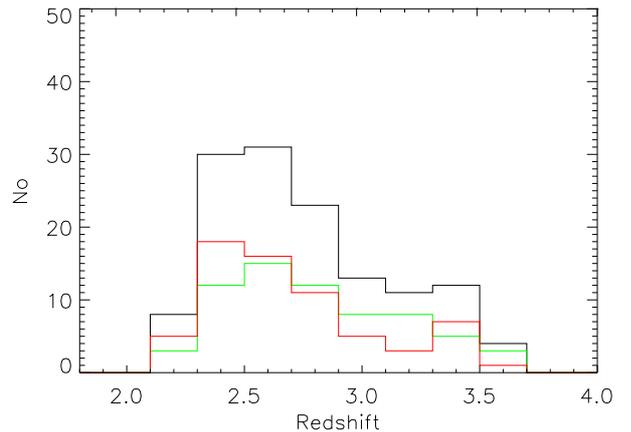}
\caption{Distribution of spectroscopic redshifts for the entire sample (black line), the \lbgl\ sub-sample (green line) and \lbgn\ sub-sample (red line).
 }
\label{fig:red}
\end{figure}
 We  found 
about 450 U-dropouts in the entire GOODS-South.
We then cross correlated this list  with the  
spectroscopic observations available in the area.  
Spectra  of galaxies in the GOODS-South  field were obtained  
by several  different observational  campaigns:
the  two largest ones were conducted by ESO, with 
FORS2 (Vanzella et al. 2006, 2008, 2009) and VIMOS (Popesso et al. 2009). 
Other redshifts were obtained  with the VLT by GMASS (Kurk et al. 2009), 
by the HST/ACS/G880L grism survey  PEARS (Straughn et al. 2008) and the similar
 GRAPES project (Pirzkal et al. 2004, Malhotra et al. 2005). 
In total there are 130 U--dropout galaxies with confirmed redshift.
\\
As in P07 we define $LBG_L$ the Lyman break galaxies that show 
the Ly$\alpha$ line in emission, regardless of its strength. We measured the equivalent width (EW) of the \lya\ line from the 1D spectra, using the IRAF task splot, by marking two continuum points around
        the line to be measured. The  linear  continuum  is  subtracted
        and  the  flux  is  determined by simply summing over the pixels (no fitting of the line profile is performed).
The sample of \lbgl\ contains 66 galaxies of which 24 have rest-frame \lya\ Equivalent Width $EW>$ 20 \AA\
and, therefore, would be selected as LAEs in a narrow band survey. The rest of the \lbgl\ 
have a line emission too faint to pass that selection.
Lyman break galaxies  with no sign of emission 
or with a clear absorption are  called  $LBG_N$ and are also 66 (note  that  two galaxies show both a clear  emission line and an absorption and are included in both samples).
The two sub-samples are not biased in terms of V-band magnitude, 
or absolute rest-frame luminosity at 1400\AA\ (see first entry of Table 1). 
This indicates that, for this sample, the redshift 
confirmation of LBGs does not depend sensibly on the presence of 
emission lines 
(see also Popesso et al. 2009).
In Figure 1 we plot the redshift distribution of our final sample of galaxies, as well as for the two sub-samples separately. The overall  redshift distribution of our  LBGs, and in particular the partial lack of objects around z$\sim 3$,  is shaped by  the  spectroscopic completeness level of the GOODS catalog (plotted in Figure 6 of Popesso et al. 2009). This spectroscopic  catalog is actually formed by a compilation of many different spectroscopic surveys (as discussed earlier), each with its own selection criteria and efficiency: the completeness level is therefore a complicated function of redshift and it has a local minimum around z$\sim 3$.
For the purpose of the present work, we just note that the two sub-samples have a consistent redshift distribution.  

\section{SED fitting}
\begin{figure*}
\includegraphics[width=16cm,clip=]{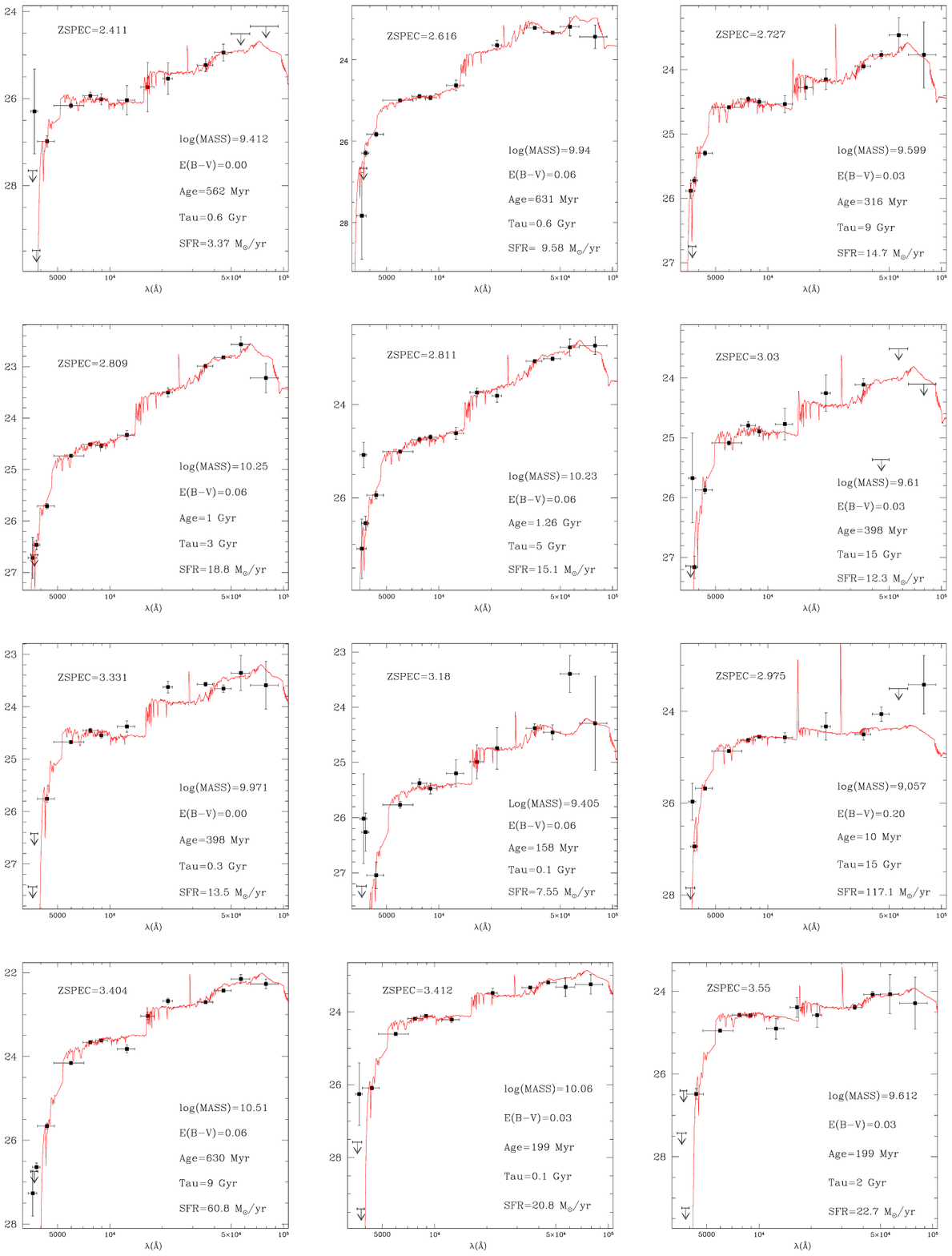}
\caption{Examples of the 14-bands SED fitting for some of the galaxies, in order of increasing redshift from z$\sim 2.4$ (upper left) to z$\sim 3.5$ (lower right). In each panel we plot the photometry in the 14 bands (U to 8 $\mu m$) in black with relative errors, and in red the best fit solution. The two emission lines that are visible in many SEDs are the [OII] and H$\alpha$ emission respectively, which are proportional to the SFR.  The best fit parameters for  each galaxy and the spectroscopic redshift are also reported.} 
\label{fig:sed}
\end{figure*}

The main physical properties of the galaxies such as total stellar mass, continuum-based star formation rate, stellar age, dust extinction E(B-V) and so on, were obtained through a spectral fitting technique which has been developed in previous papers (Fontana et al. 2003, F06), and is similar to those adopted by other groups in the literature (e.g. Dickinson et al. 2003; Drory et al. 2004). 
Briefly,   a grid of spectral templates is computed from
 standard spectral synthesis models, and the expected magnitudes in our 
filter set are calculated. The derived template library is compared with the 
available photometry and the best-fit model template is adopted according
 to a $\chi^2$ minimization. During the fitting process, the redshift 
is fixed to its spectroscopic value. The physical parameters associated with 
each galaxy are obtained from the best-fit template up to 5.5 $\mu $m
 rest-frame. This analysis assumes that the overall galaxy SED can be 
represented as a purely stellar SED, extincted by a single attenuation law, 
and that the relevant E(B-V) and basic stellar parameters 
(mostly age and star formation history, but also metallicity) can be 
simultaneously recovered with a multi-wavelength fit. We note that parameter 
degeneracies cannot be completely removed, especially at high redshift, even 
if our filter sets samples quite evenly the overall wavelength range involved. 
Previous studies (Shapley et al. 2003, 2001; Papovich et al. 2001) 
demonstrated that stellar masses are well constrained,
 while for other parameters 
the uncertainties become larger, especially at high redshifts, 
due to the SFR-age-metallicity degeneracies. 
\\
In our analysis, we estimated SFR, AGE Mass, E(B-V) and metallicity 
using Charlot \& Bruzual (2009 in preparation, hereafter CB09) synthetic models, fitting the whole 14 bands 
of photometry (namely U$_{3.5}$ and U$_{3.8}$ from the WFI on the ESO 2.2m telescope, U$_{VIMOS}$, ACS/HST B, V, i and z, VLT/ISAAC J, H and K, SPITZER/IRAC 3.5, 4.5, 5.8 and 8$\mu m$).
Note that the catalog was produced using a specific software for the accurate "PSF-matching'' of space and ground based images of different resolution and depth,  named ConvPhot 
(De Santis et al. 2007).
We assumed a Salpeter IMF and we parametrize the star formation 
histories with a variety of exponentially 
declining laws (of timescales $\tau$ranging from 0.1 to 15 Gyr), 
metallicities (from $ Z = 0.02 {Z}_\odot$ to $Z = 2.5 {Z}_\odot$) 
and dust extinctions ( 0 $<$ E(B-V) $<$ 1.1, using a Calzetti extinction curve). 
Details are given in Table 1 of Fontana et al. (2004), in Fontana et al. (2006) and in Grazian et al. (2006). The age of each object is constrained to be less than the age of the Universe at the relevant redshift.  We also adopted 
a minimum age of 10 Myrs. We are aware that exponential star formation histories may not be the correct choice in some cases, but allowing $\tau$ to vary we can actually reproduce both an instantaneous burst (when $\tau$ is very small) and a constant star formation rate (when $\tau$ is $>>$ than the age of the galaxy.)
\\
At variance with our previous works (P07, P09) we use here  
the new version of the stellar population model described 
in detail in Charlot \& Bruzual (2009). 
The main improvement over the previous  Bruzual \& Charlot 
(2003, BC03) model is an updated  treatment of   the 
emission from the stars that are in the thermally-pulsing asymptotic 
giant branch (TP-AGB) phase of stellar evolution 
from the models of Marigo \& Girardi (2007). 
 As shown first by Maraston (2005, M05) and 
Maraston et al. (2006), TP-AGB stars are 
especially important in galaxies whose spectra are
 dominated by emission from these intermediate age stars. 
\\
As already shown by several authors (e.g. Marchesini et al. 2009,
Eminian et al. 2008, Lamareille et al. 2008) 
the CB09 models give masses that are, on average, lower than the BC03 ones.
In particular,  the recent study by Salimbeni et al. (2009) 
on a sample that partially overlaps with the present one showed that  the 
average mass difference in their highest redshift bin (which includes the range z $\sim$ 2.5-3.5, see their figure 4) is around  $\sim 30$\%,  with a slight dependence on  mass. 
\\
Examples of SED fitting for some of the galaxies, spanning the entire redshift range, are shown in Figure  \ref{fig:sed}.
While [OII] and H$\alpha$ emission are included in the models, with strenght proportional to the SFR, according to the Kennicutt (1998) relations, the \lya\ and [OIII] emission line are more difficult to model.
In particular the \lya\ line could contaminate the photometry of the B or V-band (depending on the redshift) and therefore change the result of the SED fitting.
To check this, we have performed a second fit, excluding the band that contains the emission line. 
Alternatively we estimate the \lya\ line contribution
to the corresponding broad band flux (B-band for $z < 2.9$ and V band for the rest) following eq. (2) of Papovich et al. (2001) and perform a third  fit including the full set of 14 bands. This method is actually somewhat more uncertain, since the EW are estimated from the 1-dimensional spectra, while the 
broad band magnitudes  are measured from aperture photometry, and we have no information on the relative spatial  distribution of the line emitting gas and the stars in each individual object. 
In any case both fits give results that are entirely consistent with the initial one:  this indicates that the \lya\ line does not influence significantly the SED-fitting outcome, also because none of the LBG is a strong emitter/absorber. It  also confirms the  solidity of the fit: the  results do not depend on the individual band, thanks  to the large number of bands that we are using in the fits. 
Finally we also checked if the [OIII]5007 line contribution could somehow change the results of the SED-fit: the [OIII] line is harder to model, since its strenght is known to vary considerably for a given H$\alpha$ flux (e.g. Moustakas et al. 2006), and this precludes its suitability as a  SF indicator. If we include it in the SED fitting  assuming a
 mean [OIII] flux as inferred in local star-burst galaxies 
corresponding to a ratio f([OIII])/f([OII])=0.32, the output of the fit do 
not change.

\section{Physical properties: results}
\subsection{Total stellar masses, ages, dust extinction}
\begin{figure*}
\includegraphics[width=18cm,clip=]{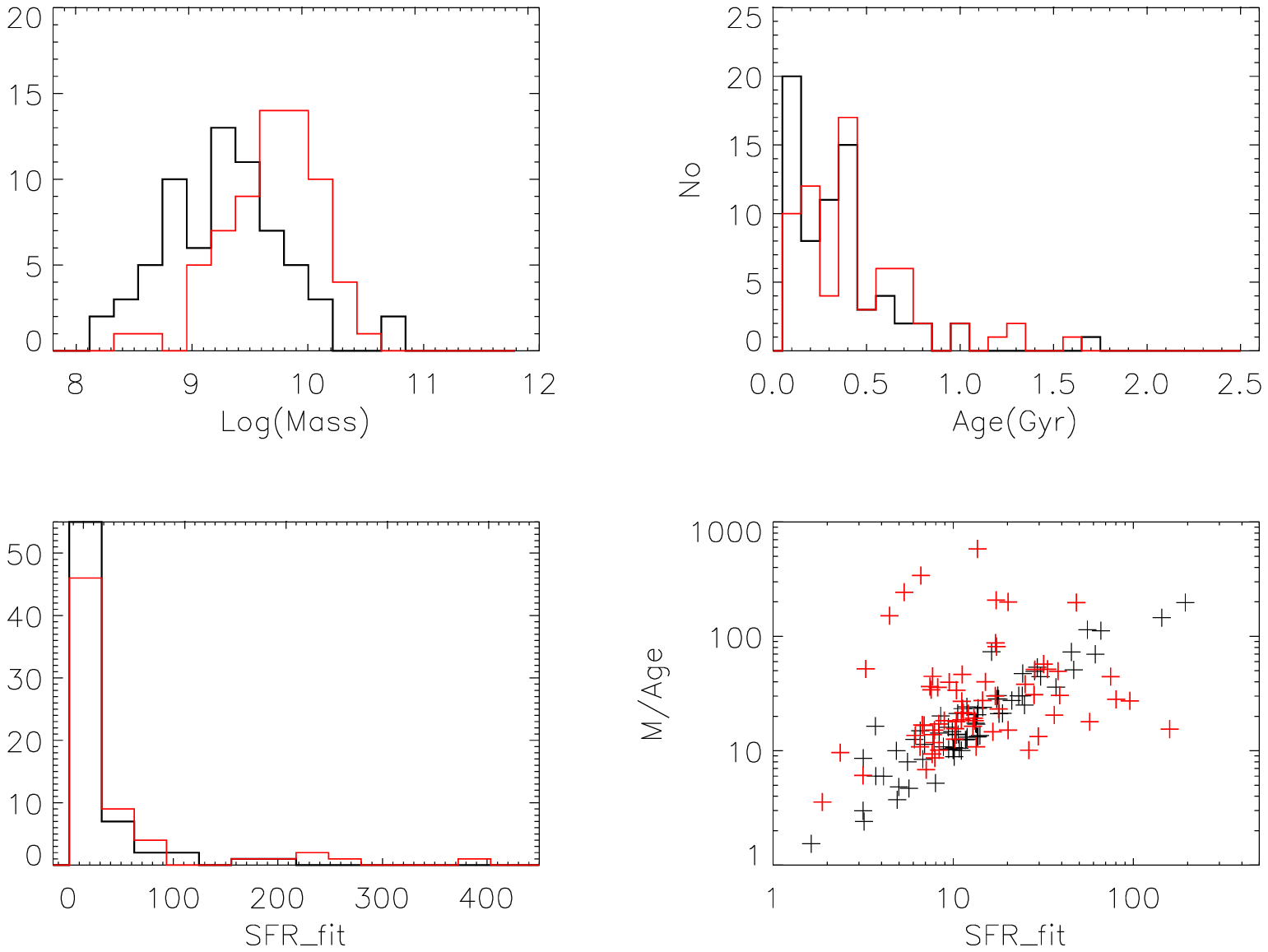}
\caption{Left: the distribution of stellar masses for \lbgl (black histogram) and \lbgn (red histogram). Right: the distribution of best fit galaxy ages. }
\label{fig:fig1}
\end{figure*}
We present here the main results for stellar masses and ages of our sample.
The values are those obtained with the CB09 models. 
For reference and 
comparison to previous work  we also 
report the relative values obtained with the BC03 (between  parenthesis). 
The total stellar masses that are found  
span a range $10^9 -10^{11} M_\odot$ , with a median 
value of  $5.3 (7.5) \times 10^9 M_\odot$. These values are similar to those found by Shapley et al. 2003 (which were obtained with the even older BC96 models).
The median age of our sample  is 320 (500) Myrs, 
very similar to the Shapley et al. value,
while  the E(B-V) best fit parameter is smaller 0.06(0.06) vs 0.15.
The discrepancies in the best fit E(B-V)  raise in part  
from the different model of star formation history adopted in the SED fitting (Shapley et al. use a constant star formation history), and in part  from the slightly different  color  selection criteria adopted in this paper, compared to 
previous works (see also end of this section).
\\
Following the approach used in  P07
we compare the properties of $LBG_L$ to those of $LBG_N$.
We find that 
the masses of $LBG_N$ are larger than the masses of $LBG_L$ and the galaxies 
without emission line are somewhat older.
The median  values with relative uncertainties, 
are reported in Table 1 and the distributions are shown in Figure \ref{fig:fig1}. 
To assess the differences between the two samples in the physical parameters
and their statistical significance,
we perform a non-parametric Kolmogorov-Smirnov (K-S) test on the age and mass distributions. We find
that the  difference is indeed highly significant for the 
mass distribution (P=0.000), while only weakly significant   
for the age distributions (P=0.136).
\\
The difference between the two samples at z$\sim 3$  
is much less pronounced than  at higher redshift: 
at z$\sim$4 the masses of $LBG_N$ were 5 times larger than the emitters, while in the present work, 
 the  difference is  only around a factor of 2.
Similarly, for the ages, the difference was larger  than a factor of 2 at 
higher z and now only around 25\% , with very low significance. Moreover, at variance with the previous work 
the age range spanned by \lbgl and \lbgn is basically equal.
\\
In P09 we found that amongst the \lbgl\ at redshift 3 to 6 
there was a significant  lack of massive galaxies with high EW. In other words,
although there was no clear  correlation between EW and  stellar mass, all 
galaxies with $EW_0 > 80 \AA$ (9 out of a sample of 70) had 
stellar masses equal or below the median mass.
There are only 3 galaxies with $EW_0 > 80\AA$ in the present sample 
and their masses are indeed all lower that the median mass.
Therefore the effect might be present, but is definitely much  less pronounced 
than at higher redshift. 
\\
The E(B-V) derived from the SED fit are small, with a median value of 0.06$\pm 0.012$ and 0.03$\pm 0.009$ for \lbgn\ and \lbgl respectively. In accordance with previous results at higher redshift (P07) and with Shapley et al. (2003),  we find that \lbgl\ tend to show less  dust extinction: the difference between the sub-sample is small but significant(PKS=0.012). To confirm this result, we  also evaluated the continuum  UV-slope from broad band  V and I data assuming that the continuum can be represented by a simple power law, of the form $f_{\lambda} \propto \lambda^{\beta}$: the median values for the two sub-sample are $\beta=-1.15\pm0.11 $ for \lbgn\ and $-1.33\pm0.12$ for \lbgl. Again, this difference indicates that there is  a diversity in dust extinction between the two sub-samples. We note that, although the trend is similar to that found by Shapley et al. (2003), they  report slightly redder slopes for their LBGs for a given EW, ranging from -0.73 for the absorbers to -1.1 for the strongest emitters (but their beta values are evaluated from the G-R colors, while our slopes are from V and I band data). On the other hand our values are similar to that reported by Hathi et al. (2008) for z$\sim 3 $ LBGs, which is  $\beta =-1.1\pm 0.2$ and by Adelberger \& Steidel (2000), which is  $\beta =-1.5\pm 0.4$.  
\\
We finally report  a significant, although very  scattered correlation 
between the total stellar mass and the extinction, in the sense that the most massive galaxies tend to show more dust.

\subsection{SFR from UV continuum and SED fitting}

\begin{figure}
\includegraphics[width=11cm]{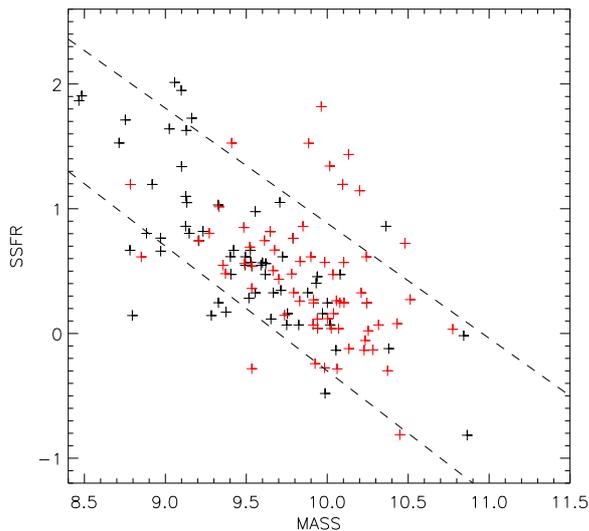}
\caption{The SSFR vs stellar mass. The two dashed lines indicate constant values of SFR of 100 and 5 $M_{\odot} yr^{-1}$. Red symbols are \lbgn\ and black ones are \lbgl.}
\label{fig:ssfr1}
\end{figure}

There are different ways of estimating the star formation rates (SFR) in LBGs.
First the  SFR can be estimated
from the  UV continuum luminosity using Kennicut (1998):
$SFR_{UV}= 1.4 \times 10^{-28}  L_\nu  M_\odot yr^{-1} $, where  $L_\nu$
 is the luminosity at rest-frame 1400 \AA\
in units of $erg s^{-1} hz^{-1}$.
This relation  assumes a $10^8$ years timescale for a galaxy to reach
the full UV luminosity, so for the youngest objects the conversion
could underestimate the true SFR. The UV emission is very sensitive to the 
presence of dust and can be attenuated even by small amounts: 
we correct for this using the slope of the UV rest-frame continuum  
determined from the V and I band data and assuming a Calzetti law. 
\\
A second value is given by the SED fit output $SFR_{SED}$. This 
clearly depends on the model assumed for 
the star formation history,
in our case the exponentially declining star formation rate
with e-folding time $\tau$.
\\
The median values for the 2 samples obtained for the SFRs in both ways  
are reported in Table 1. 
We find no significant differences between \lbgl\ and \lbgn\ in agreement  with P07 
for z$\sim $ 4 sources. 
\\
A commonly used measure to assess the importance of the current episode of star formation to the buildup of the stellar mass in galaxies is the specific  star formation rate (SSFR) i.e. the star formation rate per unit mass.
It is well known that more massive galaxies have lower SSFR (e.g. Erb et al. 2006) and that at a fixed stellar mass, the SSFR declines with increasing redshift (e.g. Reddy et al. 2006 , Feulner et al. 2005).
In Figure \ref{fig:ssfr1} we plot the SSFR vs total stellar mass: the two quantities are anti-correlated with  high probability (a Spearman Rank test gives a correlation  coefficient of -0.6). In practice, galaxies with lower stellar mass are assembling a much higher fraction of their mass with the current star formation episode, compared to more massive objects.
\begin{table}
\caption[]{Average properties of \lbgl and \lbgn derived from the SED fitting
with the Charlot \& Bruzual (2009) models}
\begin{tabular}{llll}
\hline
\hline
Property & $LBG_L$ & $LBG_N$  & P(K-S)  \\
\hline
$N_{gal}$              &  66  & 66 & \\
\hline
L(1400) (cgs)& $5.9\pm0.6 \times 10^{28}$ &  $6.0\pm0.6 \times10^{28}$ &0.993 \\
$SFR_{SED}$ ($M_\odot yr^{-1}$)   & 13.2 $\pm 4.2$ & 16.0$\pm 11$ & 0.367 \\
$SFR_{UV}$ ($M_\odot yr^{-1}$)    & 15.6$\pm 1.9$   & 12.9$\pm 3.4$  & 0.414 \\
Age (Myrs)             & 250$\pm$35   & 316$\pm$40 & 0.136\\
E(B-V)                 & 0.03$\pm0.009$   & 0.06$\pm0.012$ & 0.012\\
Mass ($M_\odot$)& 3.4$\pm 1.5 \times 10^9$& 8.7$\pm 1.2 \times 10^9$ & 0.000 \\
\hline
\hline
\end{tabular}
\label{popgal}
\end{table}
\subsection{MIPS  detection and IR-determined SFR}
At redshift $\sim 2.4$ the mid-IR  (5--8.5 $\mu$m) features associated with PAH emission, which are ubiquitous in local and z$\sim$1 star-forming galaxies, are shifted into the MIPS 24$\mu$m filter.
We searched  for counterparts of our LBGs in the GOODS-MUSIC 24$\mu$m MIPS catalog (see Santini et al. 2009 for details).
Note that galaxies with mid-IR excess such that the  MIPS flux is probably 
due to AGN activity rather than star formation (see also Fiore et al. 2008 for the proper definition) are flagged as AGNs in the revised version of the 
GOODS-MUSIC catalog that we are using and therefore were  excluded 
from our initial sample selection (see also Section 2).
Therefore we assume that the 24$\mu m$ flux is due to star formation.
\\
We found a total of 39 objects with clear detections.
Of these, 12 are \lbgl\ and 27 are \lbgn\ indicating that  
 the detection rate of \lbgn\ is more than double, compared to the line 
emitters. 
As expected the subset of MIPS detected sources is on average brighter 
(in z--band) and  at a slightly lower redshift than the entire sample. 
The  MIPS detected sources all belong to the 
most massive end of the stellar mass distribution, while other properties 
such as age and morphology (see later)  are in the average range.
In particular the mean log(Mass) of MIPS detected galaxies is 0.48 dex 
higher than the rest of the sample, and this can explain the higher detection rate for \lbgn. 
\\
Reddy et al. (2006) found similar results for  optically selected 
($UGR$)  star-forming galaxies at z$\sim$ 2:
the age distributions of 24$\mu m $ detected and non-detected 
 galaxies are similar, while the  mass distributions 
are offset such that undetected galaxies have an average $log(M)$ of  
 0.4 dex lower than for  24$\mu$  detected galaxies.
The MIPS detected galaxies have weaker \lya\ emission and 
stronger interstellar absorption lines, consistent with our results.
\\
Following the approach outlined in
Santini et al. we derived the total IR luminosity  and then the 
MIPS-Inferred total SFR. As noted by Reddy et al. 2006, the conversion from the 24$\mu m$ flux to total IR luminosity and then to unobscured SFR are subject to many uncertainties. 
For our galaxies, accurate spectroscopic redshifts are
 known and this gives the
advantage of knowing precisely the location of the PAH features
and their contribution to the 24 micron flux. Nevertheless we point out that there 
are still many sources of
uncertainties in the conversion from 24 micron flux to total IR
luminosity, due to the modeling of the PAH features, which depend on the
galaxy metallicity, galaxy environment, galaxy size, star formation
history, and on the ionizing radiation field (Calzetti et al. 2007).

\begin{figure}
\includegraphics[width=11cm,clip=]{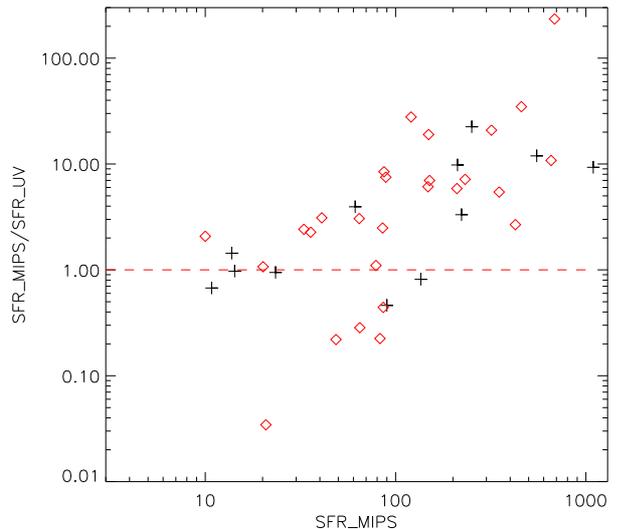}
\caption{Ratio of MIPS derived SFR to SED fitting values, versus the MIPS-SFR.
Red symbols are \lbgn\ and black symbols are \lbgl. Although the rate of MIPS detection for \lbgn\ is double than that  of \lbgl\ the two sample's fall on the 
same relation.  }
\label{fig:mips1}
\end{figure}

The derived $SFR_{MIPS}$ are on average higher than both $SFR_{SED}$ and $SFR_{UV}$. We also find a clear trend that the $SFR_{MIPS}/ SFR_{SED}$  tend to be
$\sim 1$ for objects with moderate star formation rates, while becoming $>>1$ for objects with large SFR, as  shown in Figure \ref{fig:mips1}. Both \lbgl\ and \lbgn\  seem to follow this trend.
One possibility is that, in objects showing  $SFR_{MIPS} >>SFR_{SED}$, the  MIPS emission is not (entirely) due to star formation but might be still partially due to AGN contribution, even after the exclusion 
of the clear obscured AGNs. However since the trend extends also to values lower than unity 
and is present at all redshifts (see Santini et al. 2009), it 
is possible that it reflects a change in some  intrinsic physical property 
of star-forming galaxies, such as metallicity. Galaxies with sub-solar 
metallicities have lower mid-IR emission 
(at least at $8~\mu$m, Calzetti et al. 2007) and higher UV luminosity than solar ones, for a given level of SFR. The observed trend could  therefore be related to a metallicity trend. Unfortunately, a direct check of this is not feasible. Reliable metallicities cannot be inferred from broad-band SED fitting, and high resolution spectroscopy is necessary to properly distinguish between SEDs characterized by different lines and hence metallicities.

Finally the observed trend could also be due to a failure of the assumed modelling of star formation history for example in galaxies with very high star formation.

\section{Spectral properties}
\begin{figure}
\includegraphics[width=9cm]{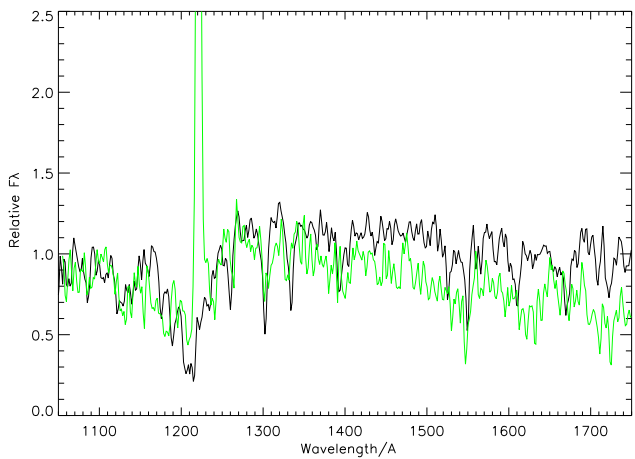}
\caption{The stacked spectra of \lbgn\ (black line)and \lbgl\ (green line). Beside the obvious difference in the \lya region, the different spectral slope and strength of the absorption lines are also clearly shown.  }
\label{fig:spectra}
\end{figure}

The majority  of the spectra were retrieved from the GOODS VIMOS  database.
Since relatively  few  spectra and redshifts were obtained from 
the other programs, we decided to make a composite spectrum using 
only  the VIMOS data. The homogeneity of the data set in terms of resolution, 
depth and calibration accuracy, makes it easier to add up the spectra, even if 
the number of sources is reduced. 
\\
The 1-D spectra are distributed in fits format from the GOODS database\footnote{http://www.eso.org/science/goods/spectroscopy/vimos.html }:  they are calibrated both in flux and wavelength.
For details on reduction and calibration see Popesso et al. (2009). The final redshift uncertainty quoted there is $\sigma_z=0.0013$, obtained from  a 
comparison between redshifts of the same objects obtained more than once with different set-ups. 
\\ 
The spectra were brought to rest-frame and co-addition
was performed adopting  a sigma--clipping procedure to eliminate 
sky-subtraction residuals, zero-count pixels due to
CCD gaps, and other spectral blemishes. After several
tests, the best results were obtained when a single 3$\sigma$
clipping iteration was adopted. On average more than
90\% of all galaxies in a given bin contribute to the stacked
spectrum at any given wavelength. 
\\
In Figure \ref{fig:spectra} we present the composite spectrum of all $LBG_L$ and compare it
to the $LBG_N$ spectrum, after normalizing the two spectra at 1280 \AA.
Besides the obvious difference of the presence of bright Ly$\alpha$ in the 
spectrum of $LBG_L$, we note that the spectrum of $LBG_N$ is 
flatter than the  spectrum of $LBG_L$, in agreement with the higher value of E(B-V) derived from the SED fit and  indicating  in \lbgn\ the presence of 
more dust than in the \lbgl. 
Furthermore, in the $LBG_N$ spectrum the 
interstellar absorption lines are more prominent than in the \lbgl\ spectrum:
this is more evident for  the low ionization lines (SII1260, OI+SiII1303, CII1334, SiII1536) 
while for the high ionization lines the difference is not as marked.
This was already observed in more details by Shapley et al. (2003), who found that the kinematic offset between the 
 Ly$\alpha$ line and the interstellar absorption lines also depends on 
Ly$\alpha$ emission strength.
A detailed analysis of the spectral stacks as Shapley et al. is not allowed by the S/N of our composite spectra. Our aim here  was only 
to check the  resulting  spectral stacks  for consistency  with previous work.

\section{X-ray emission}
The field has been observed by Chandra in the 2Msec exposure (Luo et al. 2008).
We therefore searched for signatures of X-ray emission from our sample of
 $LBG_L$ and $LBG_N$. \\
To start  we remind that, in the initial sample  selection, we 
had excluded spectroscopically confirmed 
broad line and narrow line AGNs (BLAGNs and NLAGNs 
respectively). 
The first (BLAGNs) are easily 
spotted since they usually show a broad Ly$\alpha$ line,  
with typical FWHM of several thousands km/s, clearly resolved 
even in our low resolution spectra.
The NLAGNs are recognized by the presence of high excitation emission lines 
in the spectra such as  NIV, CIV, HeII. 
Clearly in this case there could be more ambiguity, 
since the presence of these  lines depends also on the s/n of 
the spectra, on the relative line ratio and on  the spectral range observed.
Therefore  we cannot exclude that 
some NLAGNs might still be included in our sample. 
\\
Assuming that we have removed all AGNs, we search for possible 
X-ray flux associated to the LBGs, as due to star-burst emission.
We cross correlated our catalogue 
with the 2 Msec Chandra point source list by  Luo et al. (2008).
The survey covers an area of ~436 arcmin$^2$ and reaches on-axis sensitivity limits of $1.9 \times 10^{-17}$ and $1.3 \times 10^{-16} erg  s^{-1} cm^{-2}$
 for the 0.5-2.0 and 2-8 keV bands, respectively. 
Two objects (both classified as $ LBG_L$)
 are present in the Chandra 2 Msec catalog as individual detections: 
$\#$ 11006 and $\#$  8543.
They were observed 
spectroscopically by Mainieri et al. (2006) who list them as LEX 
(emission line galaxies). However from soft band  fluxes (5.37 and 3.7 
$\times 10^{-16} erg s^{-1} cm^{-2}$ respectively) and optical 
magnitude, we derive ratios of X-ray to optical 
flux that fall within the region populated by AGNs 
(e.g. Hornschemeier et al. 2001).
\\
The detection rate is comparable to Lemher et al. (2005) 
who find  7 individually detected objects out of 449 U-dropouts, while
 Laird et al. (2006) present a higher detection rate
in the HDF-N area: in their spectroscopic sample of 89 LBGs
 they find X-ray emission from 4 galaxies, or 4.5\%.
\\
We stacked  the X-ray undetected 
sources following the procedure adopted by  Laird et al. (2006). 
The final stacked image corresponds to a total 
integration time of 0.26 Gs.
As background we used the images produces by Luo et al.
 and available at the Chandra survey web site.
The flux at each source position was summed using a box cell
 of 6$\times 6$ pixels, corresponding to approximately 3$''$  a side.
This value  was chosen following  Laird et al., who determined that a radius of 
$\sim$1.5$''$ maximizes the signal to noise in stacked detection of U-dropouts.
The total counts were summed and the background 
was determined in an area of the same aperture.
The result is a detection with S/N= 4.1 in the soft band (0.5-2Kev), and a non-detection in the hard and total bands. 
The total net source counts in the soft band is 76 for 128 sources,
 implying a mean source count  of 0.59.
This is basically equal to the result of  Laird et al., who report an average 
0.58 net counts for their sample of
 277  HDF-N U dropouts in the 2 Msec observations, while  
Lemher et al. found an average 0.44 counts per source.
Assuming the stacked X-ray flux is due to star formation activity we 
determine the average unobscured SFR implied by the X-ray. 
The flux corresponding to the average count rate is 
2.3$\times 10^{-18} ergs s^{-1} cm^{-2}$.
At the mean redshift of the sample (z=2.85) this corresponds to a luminosity in the 2-8 Kev rest-frame band of 1.62$\times 10^{41} erg s^{-1}$.
We apply the conversion from X-ray flux to SFR derived by
 Nandra et al. (2002), assuming  a Salpeter IMF (as in our models).
We derive an average unobscured  $SFR_X=29.1 M_\odot yr^{-1}$, slightly 
 higher than  both $SFR_{UV}$  and $SFR_{fit}$.
\\
Most importantly, in the context of the present work, when  
 we stack the fluxes of the \lbgl\ and \lbgn\ separately we do not find any 
 significant difference.
In particular, we find a total  S/N of 2.1 for the $LBG_L$ 
and S/N=2.6 for the $LBG_N$.
Again, we run a nonparametric K-S test on the individual  
counts for each group and 
find that the two distributions are perfectly compatible with each other 
(P=0.99).
We further probe this by stacking only the strongest emitters, with an observed 
\lya\ EW$>100\AA$ and we obtain a similar result.  
Therefore we conclude that  the X-ray properties do not depend on the presence and strength of the  \lya\ emission line or, at most, are only weakly correlated. 
\section{Morphological properties}
Even for the lowest redshift objects in the sample, the ACS z-band samples the
  rest UV-emission  ($\sim 3000 \AA$ or shorter). 
The UV morphology tends to be patchy and irregular 
even for local galaxies (e.g. Gordon et al. 2004).
However, high redshift galaxies  have irregular morphologies not only in the UV, but also at longer rest-frame wavelengths (Papovich et al. 2005, Dickinson et al. 2005),  indicating that light is always dominated by the young stellar component.
The most popular explanation is that these irregular systems represent mergers (e.g., Conselice et al. 2003). 
Alternatively, it could be that the different components  observed are 
just clumps of star formation rather than different merging systems.
It is therefore  interesting to determine  if the morphology 
is connected to the other physical 
and spectral properties of the galaxies.
\\
We have analyzed the high resolution  ACS z-band morphologies of our galaxies. In Figure \ref{fig:postage} we show the postage stamps of some of the galaxies in our sample, in order of increasing redshift.
We performed  a non parametric analysis as in  Scarlata et al. (2007).
The program initially determines  the Petrosian radius and then uses it as semi-major axis of an elliptical aperture where it calculates the following parameters: asymmetry, concentration, 
Gini coefficient, and second-order moment of the brightest 20\% 
of galaxy pixels.
The Gini coefficient (G)  describes 
how uniformly the flux is distributed among galaxy pixels. 
The Gini statistics assumes values from 0 (if the galaxy light is 
homogeneously distributed among galaxy pixels) up to 1 (if all the light 
is concentrated in 1 pixel, regardless of its position in the galaxy) 
The concentration $C=5 log(r80/r20)$  with r80 and r20 the radii including 80\% and 20\% of the total galaxy light, respectively, quantifies the central density of the galaxy light distribution. $M_{20}$ is  the second-order moment of the brightest 20\% of the galaxy flux (Lotz et al. 2004). For centrally concentrated objects, $M_{20}$ correlates with the concentration C; however, $M_{20}$ is also sensitive to bright off-centered knots of light
The asymmetry (A) quantifies the degree of rotational symmetry of the 
light distribution. A is measured by calculating the normalized difference between the galaxy image and the image rotated by 180°. A correction for background noise is also applied (as in Lotz et al. 2004). 
Finally we also use the  ellipticity ($\epsilon$) of the light distribution, as measured by SExtractor (Bertin \& Arnouts 1996 ).
\\
As shown in Scarlata et al. (2007), this non-parametric method is very efficient in quantifying galaxy morphology. 
The quantities above provide complementary, but also redundant, information on galaxy structure. For a complete description and also the uncertainties on the parameters see Scarlata et al. (2007).
We only remark here that, using  the Petrosian radius 
we are not affected by distance since
surface brightness dimming does not change the shape of a galaxy
light profile and therefore it does not affect the Petrosian index.
For this reason we can compare the morphology of objects even if our sample 
 spans a relatively large redshift range.
We are aware that our method to calculate the Gini coefficient
(inside a Petrosian aperture) might introduce some compression in its
dynamic range and somewhat limit its discriminating power 
(as argued by Law et al. 2007).
In addition to the above parameters, we have visually inspected the individual 
z-band thumbnails to check for multiple peaks of emission: clearly this is a more subjective and less precise analysis.
\begin{figure*}
\includegraphics[width=16cm]{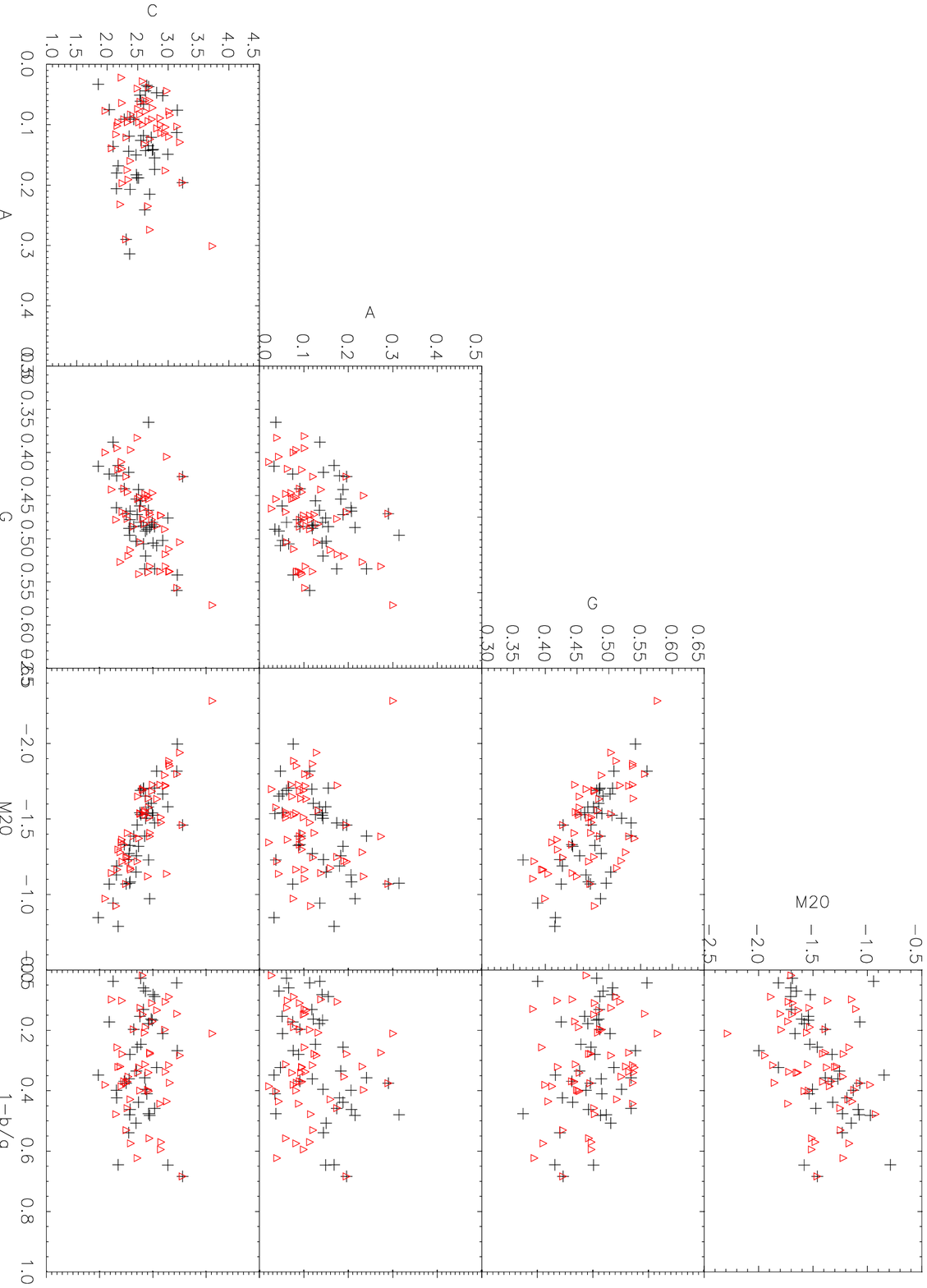}
\caption{Relations between the nonparametric diagnostics (M20, G, A, C, and $\epsilon = 1 - b/a$). The main correlations among some of the parameters, such as M20, C, and G, are clearly visible in these diagrams. In each panel black + are \lbgl\ and red triangles are \lbgn. }
\label{fig:morfo}
 \end{figure*}
\\
Separate median values for \lbgl\ and \lbgn\ are reported in Table 2.
In Figure \ref{fig:morfo} we show the G, $M_{20}$, A,  C and $\epsilon$ values 
for all galaxies
analyzed, with separate symbols and colors for  \lbgl\ and \lbgn.
\\
We find that the z$\sim$ 3 LBGs have median values G$\sim$ 0.48, $M_{20} \sim -1.46 $,   C $\sim 2.6$ and A$\sim 0.12$.  
Comparison of the absolute values to those derived by other studies 
 is not meaningful since they depend on the S/N ratio and on the choice of aperture within which they are measured (Lisker 2008).  
However our aim is a relative comparison between \lbgn\ and \lbgl: this is possible because the possible systematic errors are similar for both populations.  
\\
As for the physical parameters, a series of   Kolmogorov-Smirnov  tests was performed to assess the probability that in each case the \lbgl\ and \lbgn\ are drawn from  different distributions for all the five parameters (A, G, C,  $M_{20}$ and $\epsilon$), plus area and size.  
The probability P for each test is also reported in Table 2.
In almost  all case (with the exception of size and area)
there are no substantial differences in the morphology  of galaxies.
The morphological analysis was also repeated restricting the two sub-samples, 
\lbgn and \lbgl, to galaxies brighter than $z_{850}< 25 $, and the results do not change.
\\
This is  at variance with what suggested by Vanzella et al. (2009)
who found a higher concentration for z$\sim 4-5$ galaxies with \lya\ emission. Note however that their sample is smaller and the 
claimed  difference has less than 2$\sigma$ significance. 
Like Vanzella et al., we also find that galaxy size and area depend on line emission properties: however this might be  a consequence of the lack of bright (massive) galaxies with large EW, rather than a real dependence on line emission. If we plot the size vs EW of our galaxies 
(analogous to Figure 19 of Vanzella et al.) we find 
that galaxies with small or negative EW span the whole range of sizes  
while large EW objects tend to be very small. 
This same trend was found by Law et al. (2007) in an analysis of combined u,v,i and z band data on LBGs at z$\sim$3.  They also find a possible correlation between Ly$\alpha$ and Gini parameter, but they note that the correlation is not  well defined.
\\
\begin{table}
\caption[]{Average morphological properties of \lbgl and \lbgn  }
\begin{tabular}{llll}
\hline
\hline
Property & $LBG_L$ & $LBG_N$  & P(K-S)  \\
\hline
$N_{gal}$              &  66  & 66 & \\
\hline
Concentration      &2.55 &2.57 & 0.52\\
Gini               &0.47 &0.46 & 0.54\\
Asymmetry          &0.11 &0.097  & 0.291\\
$M_{20}$           &-1.40 &-1.41 & 0.989 \\
Ellipticity       & 0.258   &0.323 & 0.203\\
Size   & 3.89    & 4.62& 0.056\\
Area  & 154    &207 &  0.020\\
Clumpy obj. & 19/66 & 19/66 & -- \\
\hline
\hline
\end{tabular}
\label{popgal}
\end{table}
In conclusion we do not find much evidence for a strong morphological 
dependence of the emission line properties, but we find that line emitters 
tend to be small galaxies, while amongst \lbgn\  there are both small and large galaxies.
\section{Summary and discussion}
We briefly summarize the main results of this work and then 
attempt to interpret them in the context of a simple scenario.
\\
1. At z$\sim 3$ \lbgn\ are significantly more massive than \lbgl.
\\
2.The ages of  \lbgl\ and \lbgn\ are comparable. 
The median age of \lbgn\ is somewhat larger 
than the median age of \lbgl\  but at less than 3$\sigma$ level. 
\\
3. The current SFR ($SFR_{UV}$ or $SFR_{SED}$, is 
similar for both groups, as expected given the initial selection.
This is further confirmed by the stacked X-ray flux, related to unobscured star formation: no relevant differences are observed  between the two groups.
\\
4. Given that ages and SFR are similar for both \lbgn\ and \lbgl\, the larger average masses of  \lbgn\ imply that these galaxies had higher star formation in the past.
We remind that $SFR_{UV}$ is sensitive only to current star formation rate, on timescales of less than 100 Myrs, while many galaxies have ages larger than this.
\\
5. The SSFR is a strong function of Mass, and is higher for low mass objects and therefore tends to be higher for \lbgl.
\\
6. The MIPS detection rate of \lbgn\ is 2.5 times  higher than for \lbgl.
\\
7. All LBGs have small dust extinction as expected: \lbgn\ have relatively higher dust content, as inferred by the larger  E(B-V) values derived from the SED-fitting and by the steepness of the stacked spectra.
\\
8. We find no 
notable  differences in  the morphological parameters, with the exception of the smaller sizes and areas of \lbgl\ compared to \lbgn.

\subsection{A simple scenario}
The continuity and almost substantial overlap of 
physical and morphological properties of LBGs, and the few trends we have found
are in substantial agreement with the unification scheme between 
LBGs and LAEs  proposed by 
Verhamme et al. (2008). In their model, most  all LBGs have intrinsically high   
Ly$\alpha$ emission ($EW\sim 60--80 \AA$ or larger), and the observed 
variety of  Ly$\alpha$ strengths and profiles, and ultimately the fact that these galaxies are selected as LBGs or LAEs is due to variations in the content of dust and HI (see also Atek et al. 2009). The assumed geometry
is that of an expanding, spherical, homogeneous, and isothermal
shell of neutral hydrogen surrounding a central star-burst emitting
a UV continuum plus \lya\ recombination line radiation from its
associated H II region: dust  is uniformly mixed to the HI gas and the variation of these two quantities shape the profile of the \lya.
 Ultimately  the main parameter responsible for these variations may be the galaxy mass. Indeed, if the most massive galaxies  also contain more dust, as we observe given the correlation between E(B-V) and stellar mass (although with a large scatter), 
a natural consequence is that massive galaxies would show Ly$\alpha$ with smaller EW or Ly$\alpha$ in absorption. This would naturally produce 
the large mass segregation observed  between \lbgl\ and \lbgn, which is one of the strongest result of the present work. It would also explain the lack of massive galaxies with bright \lya\ emission observed here and already pointed out in P09, and similarly the lack of luminous galaxies with bright  \lya\ emission observed by other authors (e.g. Ando et al. 2007).
\\
Furthermore, in the Verhamme et al. model, no age constrains 
are derived from the presence of \lya\ up to $EW\sim 100 \AA$, as such equivalent widths can be obtained, expecially in the case of a constant star formation history, even taking into account various degrees of dust suppression. This is at variance with other models that put strong limitation of the maximum age of \lya\ emitting galaxies (e.g. Mao et al. 2007, Mori \& Umemura 2006). 
Therefore our result that \lbgl\ and \lbgn have very similar ages, and the fact that we do find \lya\ emitting galaxies with ages exceeding several hundred Myrs are also nicely in agreement with the Verhamme et al. model.
They actually predict that objects with very bright \lya\ (EW$>> 100 \AA$) should all be extremely young (less than 10-40 Myrs), but in our sample of \lbgl\ we do not have any of these large EW galaxies. Last but not least, the lack of 
significant morphological differences between \lbgl\ and \lbgn\ is also in agreement with this scenario. 
\\
Possible outlier in this unifying scheme
would be  objects that have \lya\ absorption but 
at the same time show no indications for dust extinction,  basically \lbgn\ with $E(B-V)=0$. Most of the \lbgn\ indeed have $E(B-V) \neq 0$, and only 3 of them 
seem to be dust-free galaxies with no \lya\ in emission.
Clearly the E(B-V) we estimate is subject to errors, 
so it could still be that a little dust is present:
alternatively in these cases the  $n_{HI}$ could 
be playing a major role in suppressing the \lya\ emission and/or there could be variations in the dust/HI ratio.  Indeed  
Steidel (2008) argue that the Ly$\alpha$  emission strength and apparent 
redshift is strongly affected by the presence of gas near the systematic 
velocity of the galaxy, which is often absent in lower-mass objects.
Therefore low mass objects naturally tend to show brighter 
\lya\ emission than high mass ones.
\\
To further test the above scenario, one would need to establish a solid relation between the total stellar mass of a galaxy and the neutral hydrogen content. Also the exact dependence between  the amount of dust and the stellar mass, as well as 
the variation of dust/gas ratio must be still 
explored in details.
Last but not least,  dust geometry and the way dust  and emitting sources 
are mixed might also influence  the observed \lya\ appearance.
Recently Scarlata et al. (2009) considered various possibilities for the geometry of dust around emitting sources, and found that the uniform dust screen in not able to reproduce all the observations, while a clumpy dust distribution 
does a better job, with no need to invoke differential 
extinction of \lya\ and continuum photons for most of the galaxies.
\begin{figure*}
\includegraphics[width=16cm]{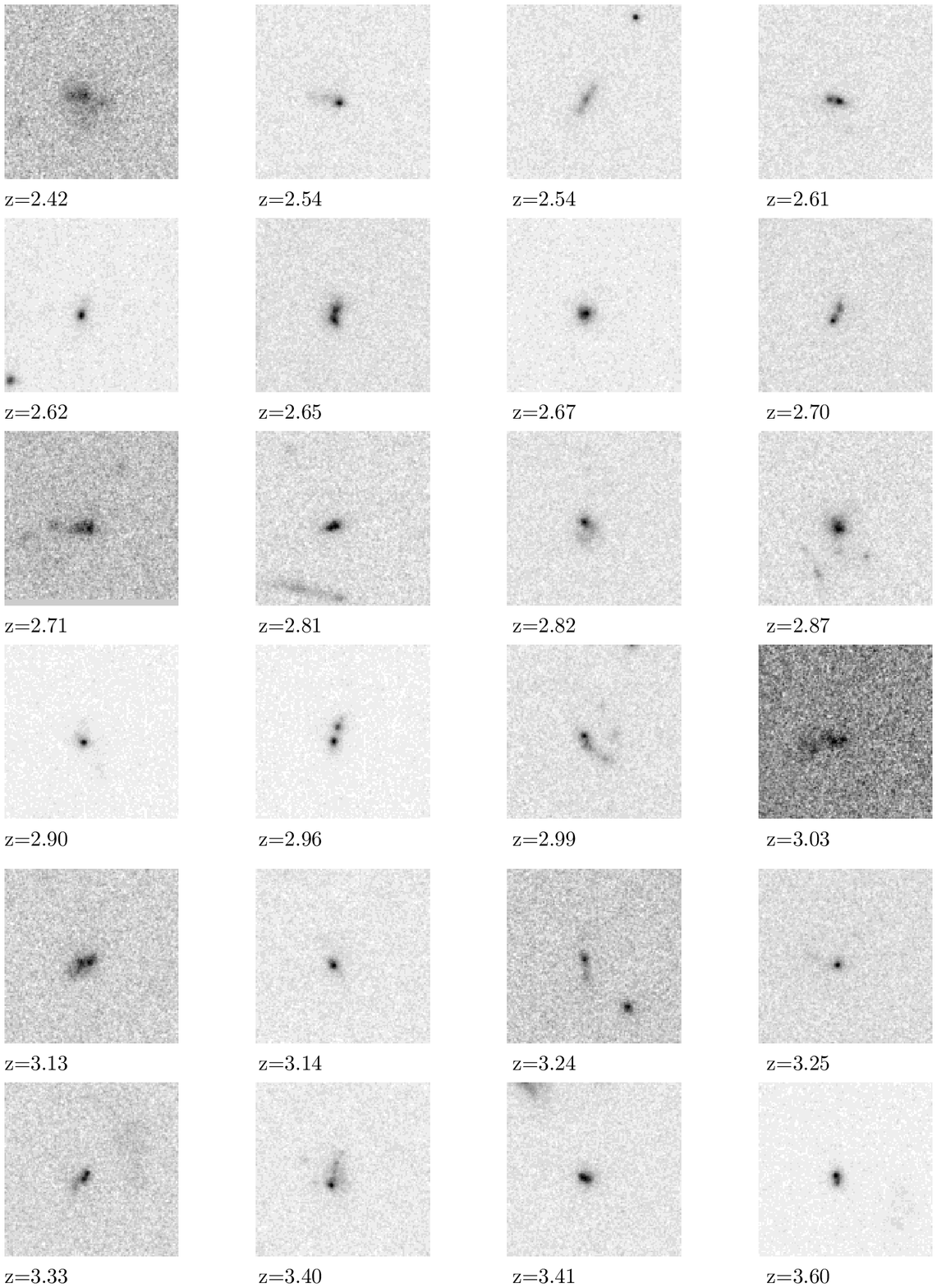}
\caption{ Postage stamps of some of our LBGs, in order of increasing redshift. Each image is taken in the z-band and is 4$'' \times 4''$ in size. }
\label{fig:postage}
 \end{figure*}
\subsection{Redshift evolution?}
Although the simple scenario described before seems to fit well the correlation found at z$\sim$ 3, as well as the absence of trends in other properties, some modification might be needed to account for the evolution in some of these properties/trends 
with redshift.
Indeed, there are strong  indications that some of the 
 correlation  between Ly$\alpha$  strength and LBGs properties
 change considerably with cosmic epoch.
First of all, as discussed in the introduction, the simple fraction of LBGs that are also LAEs, and ultimately the EW distribution of \lya\ in LBGs at the various cosmic epochs, is still subject to debate.
Most authors claim that the distribution of \lya\ strength found at $z=3$ by Shapley et al. (2001) remains valid at all redshifts  (e.g. P07, Stanway et al. 2007, Dow-Hygelund et al. 2007), although the z $\sim$ 6 population has a tail of sources with high rest-frame equivalent widths (Stanway et al. 2007). Other authors  claim instead that at very high redshift 
almost all LBGs strong \lya\ emitters  (e.g. Shimasaku et al. 2006).
Finally,  at lower redshift, Reddy et al. (2008) noted that, amongst UV selected star-forming galaxies, the redshift$\sim 3$ (LBG) population has a higher incidence of \lya\ in emission than the redshift$\sim 2$ (the so called BX) population.
\\  
In this work we found further 
indication of evolution for other properties: first of all 
at z$\sim 3$, we find no age segregation, while  at z$\sim$4,  we 
reported  that \lbgl\ were considerably much younger than \lbgn (P07). The difference in total stellar mass was also much more pronounced at z$\sim 4$ (P07), compared to this study.
\\
At z$\sim$ 3 we also find no significant evidence for morphological 
differences between \lbgn and \lbgl, but at  lower redshift Law et al. (2007),
using  a different set 
of morphological  parameters, found  possible although  not well 
defined trends.
On the other hand, at $z\sim 4$ Vanzella et al. (2009) reported a marked 
 difference in the Gini and concentration indexes of the two groups, 
with the \lbgl\ being more concentrated than the rest of the galaxies.
Similarly at $z\sim 6$ the tentative results 
of Dow-Hygelund et al. (2007) showed 
that sources with \lya\ emission are smaller, on average, than 
the i-dropout population in general.
However at variance with this study, Taniguchi et al. (2009)  recently claim that there is no difference in the morphological properties between the two populations, LAEs and LBGs at z$\sim$6.
These discrepant results indicate that disentangling the morphological properties 
of high z galaxies and comparing different samples at different redshift
is still far from trivial. In a forthcoming work we plan to analyse more in depth the morphologies of LBGs at various redshifts, using a unique set of parameters, so that a consistent  comparison can be made between the various cosmic epochs.
\\
In any case, the above results strongly suggest an evolution of some  
some fundamental property 
of galaxies, be it  the dust content (e.g. Nillson et al. 2009) or \lya\ escape fraction (e.g. Nagamine et al. 2008), that shape the appearance of star-forming galaxies: it is therefore essential that they are taken into account by the models.

\end{document}